# Specific Heat of Disordered $^3$He

H. Choi[a], J.P. Davis[a], J. Pollanen[a], N. Mulders[b], and W.P. Halperin[a]

[a]*Department of Physics and Astronomy, Northwestern University, Evanston, IL 60208, USA*
[b]*Department of Physics and Astronomy, University of Delaware, Newark, DE 19716, USA*

**Abstract.** Porous aerogel is a source of elastic scattering in superfluid $^3$He and modifies the properties of the superfluid, suppressing the transition temperature and order parameter. The specific heat jumps for the *B*-phase of superfluid $^3$He in aerogel have been measured as a function of pressure and interpreted using the homogeneous and inhomogeneous isotropic scattering models. The specific heat jumps for other *p*-wave states are estimated for comparison.

## INTRODUCTION

Porous silica aerogel imbibed with liquid $^3$He provides a source of scattering and modifies the superfluid phases. Torsional oscillator[1], NMR[2,3], and specific heat experiments[4] performed on $^3$He in aerogel show substantial suppression of the transition temperature and order parameter. Thuneberg *et al.*[5] describe a model for such a system in the framework of the Ginzburg-Landau theory, where the free energy of various superfluid $^3$He phases is expressed with different combinations of the $\beta$ coefficients[6]. As a result, the specific heat jump of $^3$He at the superfluid transition temperature can be calculated for the possible *p*-wave states of the superfluid.

According to the homogeneous isotropic scattering model (HISM), the impurities, i.e. the silica balls in the aerogel, are assumed to be distributed homogeneously in space. Then only one parameter, the quasiparticle mean free path, is needed to describe the effects of impurities in the strong scattering limit. The HISM, however, systematically underestimates the transition temperature. Sharma and Sauls introduced a second parameter in the inhomogeneous isotropic scattering model (IISM) the strand-strand correlation length $\xi_a$, to make the pressure-temperature phase diagram consistent with other superfluid properties[7]. The mean free path was estimated to be 150 nm from previous measurements of the phase diagram[1], consistent with transport measurements of spin diffusion[8] and thermal conductivity[9,10] of liquid $^3$He in aerogel. Using this mean free path at zero magnetic field, the *B*-phase is predicted to be stable[5] and experiments agree[11]. We investigated the predictions of both the HISM and IISM on the specific heat jump for *B*-phase of superfluid $^3$He in aerogel and compared the calculation with our experiment[4].

## EXPERIMENT

The sample cell used in this experiment was made of high purity silver. It was cooled by adiabatic demagnetization of $PrNi_5$ after precooling with a dilution refrigerator. The cell was thermally isolated with a cadmium heat switch; then an adiabatic heat pulse technique was used to measure the specific heat. An LCMN thermometer was calibrated with a melting curve thermometer[4,12]. The measurement was performed in the pressure range of 1 to 29 bar and the temperature range of 0.75 to 5 mK.

## RESULTS AND DISCUSSION

Fig. 1 shows the phase diagram obtained from the specific heat measurement. Both the transition temperature and the specific heat jump are suppressed in aerogel $^3$He; this is a direct manifestation of the suppression of the order parameter. We present here a detailed analysis of the specific heat jump to show that this is, in fact, the superfluid *B*-phase.

In the Ginzburg-Landau theory, the specific heat jump for a given phase, $\Delta C_p$, is calculated in terms of the $\alpha$ and $\beta$ coefficients, $\Delta C_p = (\alpha'(T_c))^2/\beta_p$, where the subscript $p$ denotes the superfluid phases of $^3$He, axial ($A$), isotropic ($B$), polar (P), and planar (PL) phases. For each of these phases, $\beta_p$ is defined as $\beta_A = \beta_{245}$, $\beta_B = \beta_{12} + \beta_{345}/3$, $\beta_P = \beta_{12} + \beta_{345}/2$, and $\beta_{PL} = \beta_{12} + \beta_{345}$, respectively.

For the HISM, the $\alpha$ and $\beta$ coefficients are modified to include the pair breaking parameter $x = \hbar v_F/2\pi k_B T\lambda$ where $\lambda$ is the mean free path[5]. The specific heat jump for the $B$-phase is calculated and compared with the experiment in the inset of Fig. 2. Within the HISM, the calculation is most consistent with the experiment for $\lambda = 160$ nm (solid curve). The same calculation is performed within the IISM by substituting $x$ with $\tilde{x} = x/(1+(\xi_a/\lambda)^2/x)$ where $\xi_a$ is the strand-strand correlation length of the aerogel[7]. For the IISM, $\lambda = 160$ nm and $\xi_a = 40$ nm best matched the experiment (dashed curve).

The relative stability of the $B$-phase is determined from the magnitude of $\beta$ coefficients. If $\beta_B < \beta_p$, the $B$-phase is stabilized over a $p$-phase and vice versa. This stability condition enters into the specific heat jump through the relation:

$$\frac{\Delta C_B - \Delta C_p}{\Delta C_B} = \frac{\beta_B^{-1} - \beta_p^{-1}}{\beta_B^{-1}} = \frac{\beta_p - \beta_B}{\beta_p}. \qquad (1)$$

According to the HISM or IISM, the $B$-phase is stable up to 30 bar. (See Fig. 2.) Despite the stability condition for the $B$-phase, a metastable $A$-phase is observed in the superfluid $^3$He in aerogel[11,13].

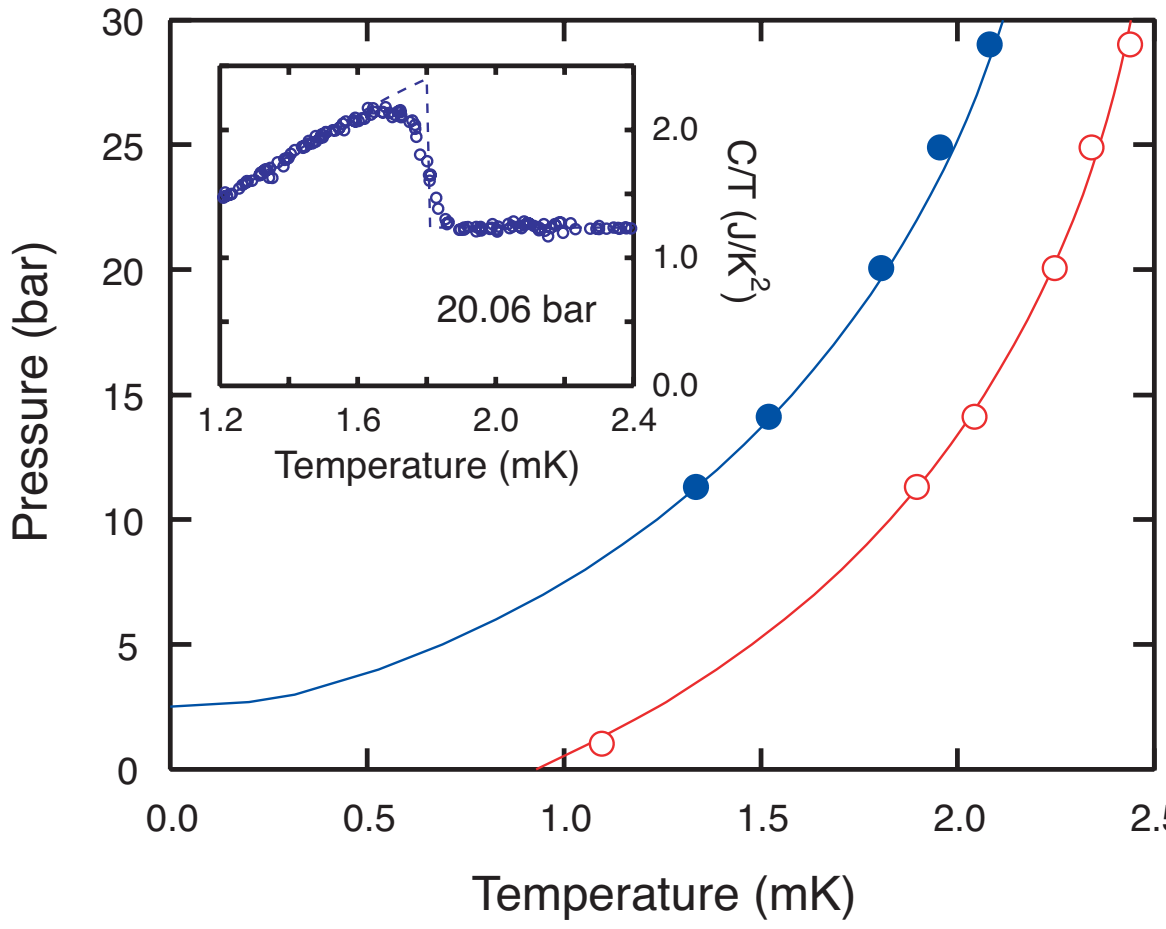

**FIGURE 1.** Phase diagram obtained from the specific heat measurement. Open circles and solid circles mark bulk superfluid transition and aerogel superfluid transition, respectively. For the aerogel superfluid transition, the IISM with $\lambda = 190$ nm and $\xi_a = 50$ nm is used to generate the curve. The inset is the $C/T$ for $^3$He in aerogel near $T_{ca}$. The specific heat jump, $\Delta C_B/C$, at $T_{ca}$ is 0.92 as opposed to $\Delta C_B/C = 1.82$ for bulk $^3$He at 20.06 bar.

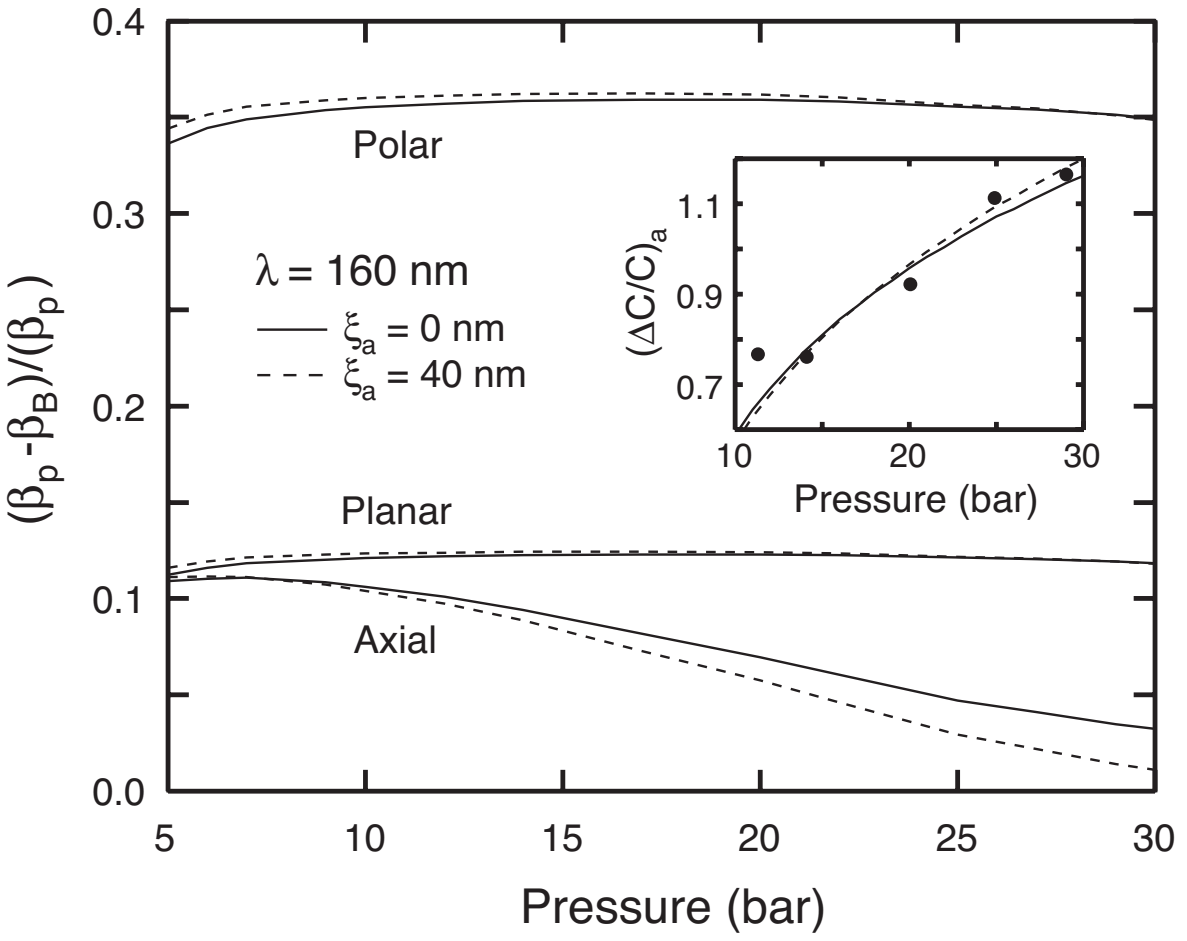

**FIGURE 2.** $(\beta_p-\beta_B)/\beta_p$ is calculated for the pressure range of 5 to 30 bar with $\lambda = 160$ nm, $\xi_a = 0$ nm and 40 nm for the HISM and IISM respectively. These parameters were obtained from the comparison between the measurement and calculation of the $B$-phase specific heat jump (inset).

The difference between the specific heat jump of the $B$-phase and the metastable phase should be observable and might be helpful in confirming its identity.

## ACKNOWLEDGMENTS

We would like to thank T.M. Haard and K. Yawata for their valuable contribution to the project and J.A. Sauls and Y. Lee for helpful discussions. This work was supported by NSF, Grant No. DMR-0244099.